# Experimental Demonstration of Stokes Space Equalization for Space Division Multiplexed Signals

F.J. Vaquero-Caballero, Gernot Goeger, Fabio Pittalà, Yabin Ye, and Idelfonso Tafur Monroy

*Abstract*— In this letter we experimentally validate, for the first time, the Stokes space algorithm (SSA) equalizer for space division multiplexing (SDM) transmission systems. We introduce the frequency domain (FD)-SSA and FD least-mean square (FD-LMS) algorithms, and evaluate their performance for different frequency offsets by computer simulations. Our simulations show that FD-SSA is insensitive to the frequency offsets, not requiring carrier frequency estimation (CFE) before equalization (pre-CFE) or carrier phase estimation (CPE) inside the loop (L-CPE). Our experimental results confirm that FD-SSA presents the same performance as FD-LMS, where the required digital signal processing (DSP) stack for FD-SSA is simpler compared to the FD-LMS.

## I. INTRODUCTION

THE achieved capacity over an optical fiber has been continuously growing at exponential rates during the previous years. The capacity growth is achieved by adopting different technologies, such as: the improvements of optical fibers with lower attenuation and dispersion; the development of Erbium-Doped Fiber Amplifiers (EDFA) [1]; coherent receivers [2], compensating of signals impairments and relaxing the requirements of the state-of-polarization (SOP) rotation; and finally the use of high spectral efficiency coding for decreasing the gap between the achieved capacity and the Shannon limit [3].

Current challenges to increase capacity are of a different nature, not motivated by technical challenges but by fundamental limitations of the optical fiber, where research efforts based on traditional methods results in diminishing returns. The research community has shown increasing interest in space division multiplexing (SDM) as the enabling technology to overcome this limitation [4]. The fundamental concept of SDM roots on the transmission of simultaneous spatial optical signals, where different approaches are currently explored.

SDM based on mode-division multiplexing (MDM), exciting multiple modes over the same fiber, requires fibers with higher numerical aperture, where few-mode fibers (FMF) or multi-mode fibers (MMF) can be considered. On the other hand, multi-core fibers (MCF) having multiple cores over the same cladding, allowing the propagation of signals over different cores with reduced crosstalk, are also studied. Hybrid approaches are also gaining popularity such as multi-core few-mode fibers (MC-FMF) [5].

MDM requires multiple-input multiple-output (MIMO) equalization as soon as mode mixing appears. As a consequence of the MDM over mode mixing systems with modal dispersion (MD), the group delay (GD) of the system is generally longer compared with single-mode fiber scenarios (SMF) [6], requiring longer filters for equalization. The digital signal processing (DSP) employed in MDM is commonly based on generalizations of polarization division multiplexing (PDM) algorithms over SMF, typically based on time domain least-mean square (TD-LMS) equalization techniques [7].

Due to the higher computational complexity of TD implementations, frequency domain (FD) equalizers are considered as an effective way to decrease computational complexity [8]. In MDM, LMS and its variants such as the Signal PSD-directed, Noise PSD-directed, or the recursive-least square (RLS) achieve different trade-offs between complexity and convergence speed [9]. Other approaches considered are channel estimation by constant amplitude zero autocorrelation (CAZAC) sequences [10] or equalization based on the constant-modulus algorithm (CMA). CMA presents advantages in terms of higher tolerance to frequency offset and phase noise [2], although CMA tendency to singularities may require a constant monitoring of its orthogonality.

The SSA was initially introduced in [11] [12] for TD and single-mode transmission. We recently proposed a FD implementation of the Stokes space algorithm (SSA) for MDM transmission system [13] [14].

In this paper we provide an experimental verification of the multi-mode equalization based on the FD-SSA algorithm in a severe scenario with high GD and mode mixing, and compare

Manuscript was written in 2016 as part of F.J. MSc Thesis. This work has been supported by the EC through H2020 project ROAM (grant no: 645361, www.roam-project.eu).

F. J. Vaquero Caballero, G. Goeger, F. Pittalà and Y. Ye are with the European Research Center, Huawei Technologies Duesseldorf GmbH, Munich 80992, Germany (e-mail: f.javier.vaquero@huawei.com; Gernot.Goeger@huawei.com; fabio.pittala@huawei.com; yeyabin@huawei.com).

I. T. Monroy is with DTU Fotonik, Department of Photonics Engineering, Technical University of Denmark, Kgs. Lyngby, DK-2800, Denmark (e-mail: idtm@fotonik.dtu.dk).

its performance to that of another singularity-free algorithm: the FD-LMS.

The structure of this paper is the following: in section II, we briefly review the theoretical foundations of FD equalization for both LMS and SSA and compare their performance with frequency offset by simulations. Section III includes the experimental setup and results and Section IV finishes with the conclusions of this paper.

## II. THEORETICAL FOUNDATIONS OF THE FD-MIMO LMS AND SSA EQUALIZATION

Figure 1 illustrates an $M \times M$ FD equalizer, based on overlap-save technique [8]. The equalization starts by concatenating 2 blocks of $N/2$ samples of the input signal ($x_v^{(n)}, y_v^{(n)}$) corresponding to the received mode $v$ ($1 \leq v \leq M$) in TD and fast Fourier-transformed (FFT) into FD:

$$X_v^{(n)} = FFT\{[x_v^{(n-1)}, x_v^{(n)}]\},$$
$$Y_v^{(n)} = FFT\{[y_v^{(n-1)}, y_v^{(n)}]\},$$
(1)

Where '$[a, b]$' operation denotes the concatenation of both vectors. The equalization is implemented in FD as a multiplication [13]:

$$\bar{X}_v^{(n)} = \sum_{j=1}^M X_j^{(n)} * H_{x_v x_j}^{(n)} + Y_j^{(n)} * H_{x_v y_j}^{(n)},$$
$$\bar{Y}_v^{(n)} = \sum_{j=1}^M X_j^{(n)} * H_{y_v x_j}^{(n)} + Y_j^{(n)} * H_{y_v y_j}^{(n)},$$
(2)

Where $'*'$ is the element-wise multiplication of vectors: $A * B = [a_1 b_1, \ldots, a_N b_N]$, $\bar{X}_v^{(n)}$ and $\bar{Y}_v^{(n)}$ are the result of the equalization and $H$ the filter coefficients. The signal is converted back into TD and the last block after equalization is saved, represented by the operator $E_{N/2+1}^N$:

$$\bar{x}_v^{(n)\prime} = IFFT\{\bar{X}_v^{(n)}\}; \quad \bar{y}_v^{(n)\prime} = IFFT\{\bar{Y}_v^{(n)}\},$$
$$\bar{x}_v^{(n)} = E_{N/2+1}^N(\bar{x}_v^{(n)\prime}); \quad \bar{y}_v^{(n)} = E_{N/2+1}^N(\bar{y}_v^{(n)\prime}),$$
(3)

The equalized signal is compared to the expected one by means of the error function: LMS or SSA. Further details of the FD equalization has been extensively covered by the research community [8] [13].

The difference between the considered equalizers lies on the domain where the error functions are calculated. LMS error function is the square difference between the expected signal and the received one in the Euclidean domain, while SSA calculates the same difference but in the transformed Stokes space. The Stokes transformation is [11]:

$$S_1(n) = |x(n)|^2 - |y(n)|^2,$$
$$S_2(n) = 2\Re\{x(n)y^*(n)\},$$
$$S_3(n) = 2\Im\{x(n)y^*(n)\},$$
(4)

Where $[S_1, S_2, S_3] \in \mathbb{R}^3$. The error coefficients $C_{1v}(n)$ and $C_{2v}(n)$ [14] are directly responsible for the filter update, the filter update rule in TD is:

$$h_{l_d m_g}^{(n+1)} = h_{l_d m_g}^{(n)} - \mu \nabla_{h_{l_d m_g}^{(n)}} f(h^{(n)}); l, m \in \{x, y\},$$
$$\nabla_{h_{l_d m_g}^{(n)}} f(h^{(n)}) = C_{pd}(n) \, m_g^{*(n)}; p = \begin{cases} 1, & l = x \\ 2, & l = y \end{cases}; m \in \{x, y\},$$
(5)

For the LMS equalizer, the error coefficients are:
$$C_{1v}(n) = 2(\bar{x}_v(n) - \hat{x}_v(n))\bar{x}_v(n),$$
$$C_{2v}(n) = 2(\bar{y}_v(n) - \hat{y}_v(n))\bar{y}_v(n),$$
(6)

For the SSA equalizer, the error coefficients are:
$$C_{1v}(n) = 4[\bar{S}_{1v}(n) - \widehat{S_{1v}}(n)]\bar{x}_v(n)$$
$$+ 4\left[(\bar{S}_{2v}(n) - \widehat{S_{2v}}(n))\right.$$
$$\left. + \sqrt{-1}(\bar{S}_{3v}(n) - \widehat{S_{3v}}(n))\right]\bar{y}_v(n),$$
$$C_{2v}(n) = 4\left[(\bar{S}_{2v}(n) - \widehat{S_{2v}}(n)) - \sqrt{-1}(\bar{S}_{3v}(n) - \widehat{S_{3v}}(n))\right]\bar{x}_v(n)$$
$$- 4[\bar{S}_{1v}(n) - \widehat{S_{1v}}(n)]\bar{y}_v(n),$$
(7)

Where '$\,\hat{}\,$' and '$\,\bar{}\,$' denotes the expected and received symbol respectively. Figure 2 illustrates both error functions. The Stokes transformation is insensitive to frequency offset and phase noise on both polarizations. As a result, SSA is insensitive to frequency offset and robust to phase noise, while FD-LMS is heavily penalized by those impairments. A CPE inside the loop (L-CPE) of the equalizer can mitigate the penalty of FD-LMS caused by frequency offsets and phase noise by compensating the estimated phase noise. The error coefficients are also multiplied by an estimation of the phase error $\hat{\varphi}_{1v}(n), \hat{\varphi}_{2v}(n)$ [7]:

$$\bar{x}_v(n) = \bar{x}_v(n)e^{-j\hat{\varphi}_{1v}(n)}$$
$$\bar{y}_v(n) = \bar{y}_v(n)e^{-j\hat{\varphi}_{2v}(n)}$$
(8)

Fig. 3 shows the required OSNR (ROSNR) of a $3 \times 3$ 34Gbaud non-return-to-zero (NRZ) dual-polarization (DP) Quadrature Phase-Shift Keying (QPSK) as a function of the normalized frequency offset. The computer simulations do not consider the bandwidth limitation which is 18GHz in the experimental setup. The flat ROSNR of the SSA reflects its insensitivity to frequency offsets. It is observable for the 64-

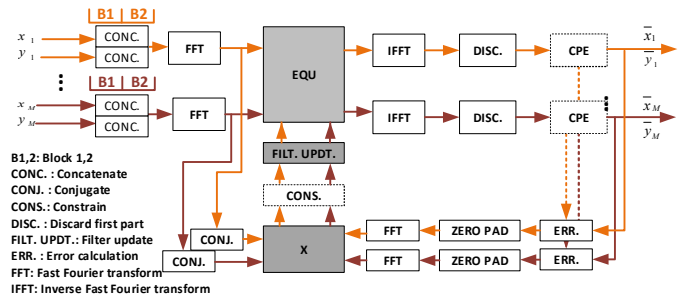

Fig. 1. Example of FD equalization

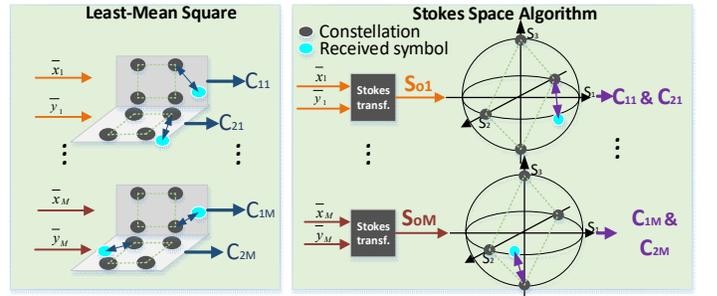

Fig. 2. LMS and SSA error functions.

taps FD-LMS scenario, a frequency offset of 1 MHz can cause 2dB OSNR penalty.

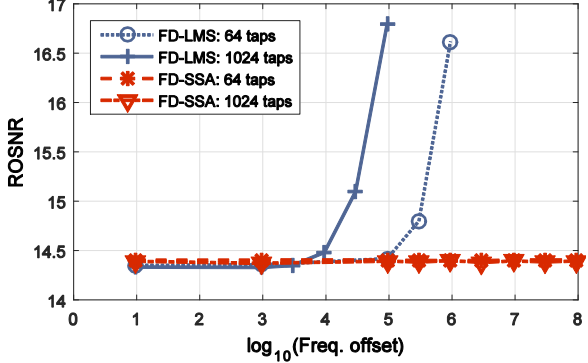

Fig. 3. Required OSNR (ROSNR) as a function of frequency offset

## III. EXPERIMENTAL RESULTS

The experimental setup is presented in Figure 4. A 34Gbaud NRZ DP-QPSK signal is generated by a digital-to-analog converter (DAC) and modulated into an optical signal by a laser at 1550nm with 100 kHz of linewidth (L). The signal is replicated by a splitter into 3 identical replicas which are de-correlated by means of different fiber segments of 5.1 and 10 meters (corresponding to 870 and 1700 symbols delay, respectively).

We aim to evaluate both equalizers in a challenging scenario, where the modulated signals are fully-mixed and multiplexed (MUX) into LP modes ($LP_{01}$,$LP_{11}$) by optical lanterns. The signal is transmitted through a 2.1 km step-index fiber with 2.1 ps/m of DMD and 21 ps/nm/km of chromatic dispersion (CD).

In the receiver side, 3 integrated coherent receivers (ICR) of 18 GHz of bandwidth, are employed for down-converting the optical signal into the electrical signal. The 3 ICRs share the same local oscillator (LO), which is different from L. Finally, 3 synchronized scopes, corresponding to a total of 12 channels, save the data for posterior offline DSP processing.

The DSP stack of our SDM receiver is similar to that for the single mode coherent DSP [2]. Both FD-SSA and FD-LMS receivers implement the same synchronization algorithm capable of finding and aligning the training data.

Then, since the implemented setup includes two independent lasers for the transmitter and receiver, the estimated frequency offset oscillates between -400 and -260 MHz (Figure 5 red), heavily penalizing the performance for FD-LMS (Figure 3). To avoid that, the FD-LMS DSP stack includes a pre-Carrier Frequency Estimation (pre-CFE) capable of reducing the frequency offset to a few MHz (Figure 5 blue) before equalization.

Though the reduction of the frequency offset enables convergence of the FD-LMS but only works for small number of taps and high OSNRs. Since the considered scenario has long impulse responses, the L-CPE is indispensable for the FD-LMS with big number of taps. A single L-CPE estimation per block is performed for the FD-LMS. The FD-SSA does not suffer from penalties in the considered frequency offset range and also FD-SSA is independent of the number of filter taps considered. Consequently, neither pre-CFE nor L-CPE is applied for the FD-SSA.

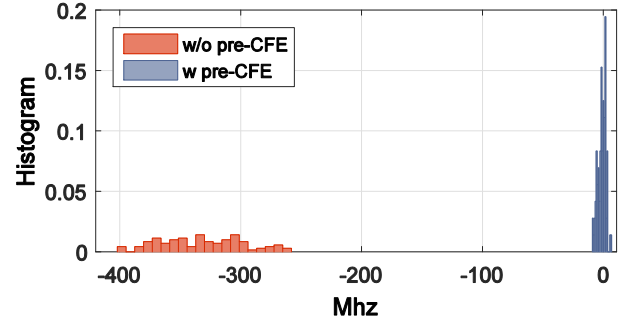

Fig. 5. Estimated frequency offset of the received signal without pre-CFE (red) and after pre-CFE compensation (blue).

After equalization both FD-LMS and FD-SSA implemented the same CFE and CPE [2] and finally the BER is calculated.

The electrical bandwidth of the system is limited to 18GHz, which is heavily penalizing the performance of our system, further penalties of the setup comes from the IQ imbalance of the transmitter which was pre-distorted to mitigate its effect. The filter coefficients are chosen to be a power of 2 for efficient FFT, being 64 taps and 1024 taps for the back to back (B2B) and 2.1Km respectively, equivalent to 32 and 512 TD equalizer taps.

Figure 6 shows the TD filter coefficients of the 2.1Km transmission: $h_{tq}$, where $q$ identifies the column number: $t, q \in$ [1,6]. Two peaks, corresponding to the $LP_{01}$ and $LP_{11}$,

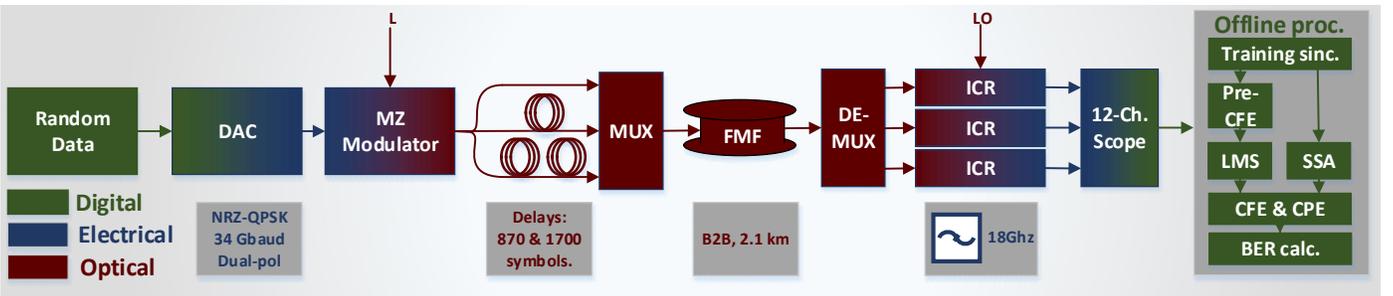

Fig. 4. Experimental setup considered.

separated by approximately 300 samples, are clearly distinguishable.

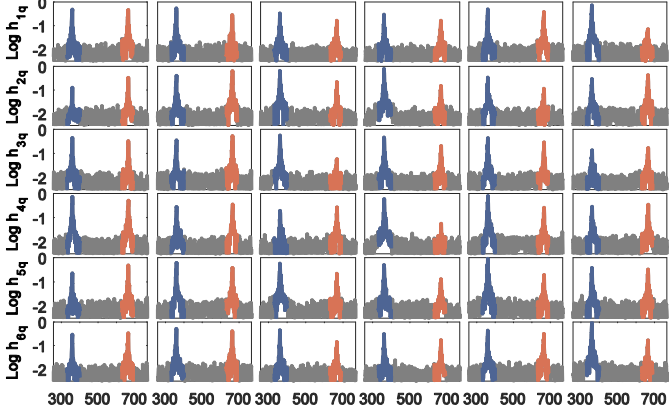

Fig. 6. $\log_{10} h$ showing $LP_{01}$ and $LP_{11}$ in different colors.

Figure 7 and 8 show the experimental results of the considered setup. For the B2B case in Figure 7, the FD-LMS with L-CPE (blue) and FD-SSA (red) achieve same performance, while the FD-LMS without L-CPE (grey) has an erratic behavior and only working for high OSNR values.

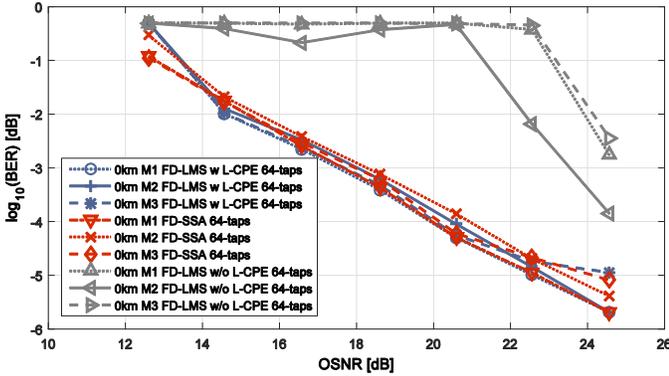

Fig. 7. Performance of the B2B transmission experiment.

For the scenario of 2.1km FMF transmission, as shown in Figure 8, FD-SSA (red) and FD-LMS with L-CPE (blue) again performs similarly. On the other hand, the FD-LMS without L-CPE (grey) is not capable of equalizing the received signal even for high OSNR due to the number of taps employed by the equalizer.

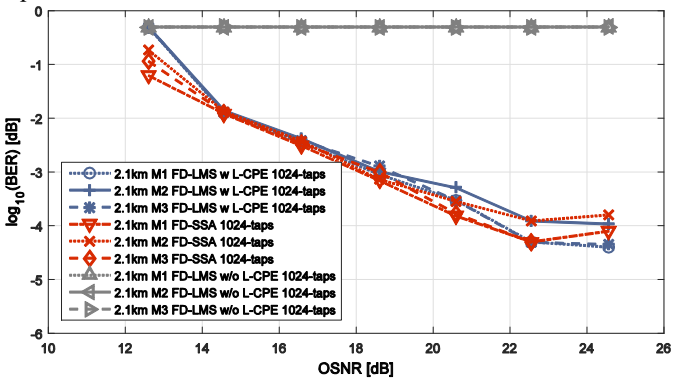

Fig. 8. Performance of the 2.1km FMF transmission experiment.

## IV. CONCLUSION

This paper presents the first experimental demonstration of the FD-SSA for SDM system and compares it to the FD-LMS. FD-SSA is an effective equalization technique in the presence of frequency offset, where the FD-LMS would require a pre-estimation of frequency offset and CPE inside the loop to achieve similar performance. The CPE inside the loop of the FD-LMS may lead to further penalties due to slower convergence and adaptation of the backward loop of the equalizer, which should be studied in later publications. In summary, FD-SSA could be the enabler equalizer for real SDM implementation by using simpler DSP stacks.